\documentclass[12pt]{article}

\usepackage{epsfig,amsmath,amssymb,verbatim,mathrsfs}

\def\beq{\begin{eqnarray}}
\def\eeq{\end{eqnarray}}
\def\bea{\begin{eqnarray}}
\def\eea{\end{eqnarray}}

\newcommand{\gsim}{\lower.7ex\hbox{$\;\stackrel{\textstyle>}{\sim}\;$}}
\newcommand{\lsim}{\lower.7ex\hbox{$\;\stackrel{\textstyle<}{\sim}\;$}}

\def\stilde{\widetilde}

\newcommand{\newc}{\newcommand}
\newc{\Nc}{N_{c}}
\newc{\CG}{C_G}
\newc{\gp}{g'}
\newc{\stopi}{\stilde t_i}
\newc{\sboti}{\stilde b_i}
\newc{\staui}{\stilde \tau_i}
\newc{\stopj}{\stilde t_j}
\newc{\sbotj}{\stilde b_j}
\newc{\stauj}{\stilde \tau_j}
\newc{\stopI}{\stilde t_1}
\newc{\stopII}{\stilde t_2}
\newc{\sbotI}{\stilde b_1}
\newc{\sbotII}{\stilde b_2}
\newc{\stauI}{\stilde \tau_1}
\newc{\stauII}{\stilde \tau_2}
\newc{\sstop}{s_{t}}
\newc{\cstop}{c_{t}}
\newc{\ssbot}{s_{b}}
\newc{\csbot}{c_{b}}
\newc{\sstau}{s_{\tau}}
\newc{\cstau}{c_{\tau}}
\newc{\Sstop}{s_{2t}}
\newc{\Cstop}{c_{2t}}
\newc{\Ssbot}{s_{2b}}
\newc{\Csbot}{c_{2b}}
\newc{\Sstau}{s_{2\tau}}
\newc{\Cstau}{c_{2\tau}}
\newc{\salpha}{s_\alpha}
\newc{\calpha}{c_\alpha}
\newc{\Calpha}{c_{2\alpha}}
\newc{\Salpha}{s_{2\alpha}}
\newc{\sbetapm}{s_{\beta_\pm}}
\newc{\cbetapm}{c_{\beta_\pm}}
\newc{\Sbetapm}{s_{2 \beta_\pm}}
\newc{\Cbetapm}{c_{2 \beta_\pm}}
\newc{\sbetaO}{s_{\beta_0}}
\newc{\cbetaO}{c_{\beta_0}}
\newc{\SbetaO}{s_{2 \beta_0}}
\newc{\CbetaO}{c_{2 \beta_0}}
\newc{\vu}{v_u}
\newc{\vd}{v_d}
\newc{\seL}{\stilde e_L}
\newc{\smuL}{\stilde \mu_L}
\newc{\seR}{\stilde e_R}
\newc{\smuR}{\stilde \mu_R}
\newc{\suL}{\stilde u_L}
\newc{\sdL}{\stilde d_L}
\newc{\suR}{\stilde u_R}
\newc{\sdR}{\stilde d_R}
\newc{\scL}{\stilde c_L}
\newc{\ssL}{\stilde s_L}
\newc{\scR}{\stilde c_R}
\newc{\ssR}{\stilde s_R}
\newc{\snue}{\stilde \nu_e}
\newc{\snumu}{\stilde \nu_\mu}
\newc{\snutau}{\stilde \nu_\tau}
\newc{\Gpm}{G^\pm}
\newc{\Hpm}{H^\pm}
\newc{\FFbS}{\overline{FF}S}
\newc{\FFbV}{\overline{FF}V}
\newc{\FSS}{F_{SS}}
\newc{\FSSS}{F_{SSS}}
\newc{\FFFS}{F_{FFS}}
\newc{\FFFbS}{F_{\overline{FF}S}}
\newc{\FSSV}{F_{SSV}}
\newc{\FVS}{F_{VS}}
\newc{\FVVS}{F_{VVS}}
\newc{\FFFV}{F_{FFV}}
\newc{\FFFbV}{F_{\overline{FF}V}}
\newc{\Fgauge}{F_{\rm gauge}}
\newc{\DRbarprime}{$\overline{\rm DR}'$ }
\newc{\DRbar}{$\overline{\rm DR}$ }
\newc{\MSbar}{$\overline{\rm MS}$ }
\newc{\Yu}{{\bf Y}_u}
\newc{\Yd}{{\bf Y}_d}
\newc{\Ye}{{\bf Y}_e}
\newc{\Au}{{\bf a}_u}
\newc{\Ad}{{\bf a}_d}
\newc{\Ae}{{\bf a}_e}
\newc{\bm}{{\bf m}}
\newc{\zhol}{Z^{\rm hol}}

\newc{\rwino}{r_{\tilde W}}
\newc{\rmu}{r_{\tilde H}}
\newc{\ra}{r_A}
\newc{\ccdot}{\!\cdot\!}

\newcommand{\phib}{\phi_{-b}}
\newcommand{\phia}{\phi_{a}}

\newcommand{\nnmb}{\nonumber}
\newcommand{\del}{\partial}

\begin{document}

\setlength{\baselineskip}{0.2in}


\begin{titlepage}
\noindent
\begin{flushright}
MCTP-06-14  \\
\end{flushright}
\vspace{1cm}

\begin{center}
  \begin{Large}
    \begin{bf}
 Holomorphic selection rules,\\ the origin of the 
 $\mu$ term, and thermal inflation
     \end{bf}
  \end{Large}
\end{center}
\vspace{0.2cm}
\begin{center}
\begin{large}
David E. Morrissey and James D. Wells \\
\end{large}
  \vspace{0.3cm}
  \begin{it}
Michigan Center for Theoretical Physics (MCTP) \\
Physics Department, University of Michigan, Ann Arbor, MI 48109

\vspace{0.1cm}
\end{it}

\end{center}

\center{\today}

\begin{abstract}

  When an abelian gauge theory with integer charges is spontaneously 
broken by the expectation value of a charge $Q$ field, there remains 
a $Z_Q$ discrete symmetry.  In a supersymmetric theory, holomorphy
adds additional constraints on the operators that can appear
in the effective superpotential.  As a result, operators with the same mass 
dimension but opposite sign charges can have very different coupling strengths.
In the present work we characterize the operator hierarchies in the effective
theory due to holomorphy, and show that there exist simple relationships
between the size of an operator and its mass dimension and charge.
Using such holomorphy-induced operator hierarchies, 
we construct a simple model with a naturally small supersymmetric 
$\mu$ term.  This model also provides a concrete realization of 
late-time thermal inflation, which has the ability to solve the gravitino 
and moduli problems of weak-scale supersymmetry.

\end{abstract}

\vspace{1cm}

\end{titlepage}

\setcounter{page}{2}



\section{Introduction}

  Consider an Abelian gauge theory with many scalar fields $\{\phi_i\}$ 
all having integer charges $\{Q_i\}$, and suppose one scalar $\phi_V$ with 
charge $Q_V$ obtains a vacuum expectation value~(VEV).
It is well-known that the low-energy effective theory
well below the VEV has a $Z_{Q_V}$ discrete symmetry~\cite{Krauss:1988zc,
Ibanez:1991hv}.
Fields can be assigned charges under the $Z_{Q_V}$
symmetry, and the effective lagrangian is built up from all
operators that are $Z_{Q_V}$-invariant combinations of the fields.

  In non-supersymmetric field theories any combination of
$\{\phi_i\}$ and $\{\phi_i^\dagger\}$ is allowed to make 
gauge-invariant operators. Therefore, $Z_{Q_V}$  is an
appropriate label for the discrete gauge symmetry of the
effective theory, since it implies no distinction between 
allowed operators that have charge $-nQ_V$ under the original 
$U(1)$ versus those that have charge $+nQ_V$ (where $n$ is positive integer).
There may, however, exist a hierarchy between operators of the same dimension
but different absolute values of their charges~\cite{Froggatt:1978nt}.

  In a supersymmetric field theory, merely stating that
$U(1)\to Z_{Q_V}$ when $\phi_V$ develops a VEV loses information.
Holomorphy of the superpotential implies that factors of 
$\langle\phi_V^\dagger\rangle$ alone cannot give rise directly to 
low-energy operators in the effective superpotential.  
As a result, the coefficients of two chiral operators 
in the low energy theory with the same mass dimension but opposite 
sign charges can be very different~\cite{Leurer:1992wg}.

  The purpose of this work is to describe the coefficient strengths 
of allowed operators in the effective superpotential as a function
of their $U(1)$ charges in the full theory.
This is the subject of Section~\ref{holom}.  In Section~\ref{muterm} 
we apply our results to construct a small value for the $\mu$ term 
in the MSSM.  In Section~\ref{therminf}, we show how this mechanism 
for the $\mu$ term naturally gives rise to a brief period of late-time 
thermal inflation, which can help to dilute overabundant or late-decaying
relics such as gravitinos or moduli.  Our conclusions are given in 
Section~\ref{concl}.  Some technical details are deferred to an Appendix.


\section{Holomorphy and discrete gauge symmetries}
\label{holom}

  Suppose a $U(1)$ gauge symmetry is spontaneously broken
in an $\mathcal{N}=1$ supersymmetric gauge theory due to
the condensation of one of more charged scalar fields.    
Even though the gauge symmetry is broken, 
the resulting effective theory will retain an invariance 
under spurious $U(1)$ transformations where the VEVs
transform as well.  The non-spurious residual discrete symmetry present
in the effective theory is the subgroup of the spurious $U(1)$ 
that leaves the VEVs invariant.  We shall make use of this spurious
symmetry in discussing the additional selection rules for 
superpotential operators due to holomorphy.  To begin, 
we discuss the case of a single VEV.  Afterwards, we generalize
to the more complicated case of two or more VEVs.

\subsection{One VEV: supersymmetric $\zhol_{Q_V}$ discrete symmetry}

  A supersymmetric field theory can remain supersymmetric upon the
condensation of a single charged field $\phi_V$ if there is a 
Fayet-Iliopoulos~(FI) term in the $D$-term potential with sign 
opposite to that of $Q_V$:
\beq
D=Q_V|\phi_V|^2-\xi + \cdots~~({\rm assuming~}Q_V,\xi>0).
\label{qvxi}
\eeq
For $\langle\phi_V\rangle=\sqrt{\xi/Q_V}$, the gauge symmetry
is broken but supersymmetry need not be.  We assume there are
no $F$-terms that would break supersymmetry for a non-zero 
$\langle\phi_V\rangle$.  We call the low-scale symmetry group in 
this case $\zhol_{Q_V}$, and the symmetry breaking path is
\beq
U(1)\stackrel{\langle\phi_V\rangle}\longrightarrow \zhol_{Q_V}.
\eeq

The leading operators allowed in the effective theory superpotential
arise in three ways:\footnote{
See the appendix for a more detailed discussion.} 
\begin{enumerate}
\item {\it Holomorphic insertions.}
$\phi_V \to \langle\phi_V\rangle$ in the superpotential of the 
full theory. For example,
\beq
W_{full} \supseteq \frac{1}{M_*^{a+d-3}}\,\phi_V^a\,\mathcal{O}^{(d)}
({\phi_i})
\longrightarrow 
\frac{\langle \phi_V\rangle^a}{M_*^a}\frac{1}{M_{*}^{d-3}}\,
\mathcal{O}^{(d)}({\phi_i})\subseteq W_{eff},
\eeq
where $\mathcal{O}^{(d)}({\phi_i})$ is an operator of mass 
dimension $d$ composed of light fields $\{\phi_i\}$, and $M_{*}$
denotes the ultraviolet cutoff of the theory.  This mechanism
generates insertions of $\langle \phi_V\rangle/M_*$ (but not
$\langle \phi^\dagger_V\rangle/M_*$), and therefore only operators 
$\mathcal{O}^{(d)}({\phi_i})$ with total charge equal to 
$0,-Q_V,-2Q_V,\ldots$ in the full theory can be generated in this way.

\item {\it Inverse-holomorphic insertions.} 
These arise from integrating out heavy fields whose masses derive from
$\langle\phi_V\rangle$.  Mass terms for the superfields in the original
superpotential are analytic in the VEVs leading to new operators
suppressed by $1/\langle\phi_V\rangle$ when these massive
fields are integrated out.
As an example, consider a theory with the superpotential
\beq
W=\lambda_1\,\phi_4\,\phi_{-2}^2+\lambda_2\,\phi_{-2}\,\phi_1^2,
\eeq
where the field $\phi_q$ has charge $q$.  If $\phi_4$ obtains a VEV,
$\phi_{-2}$ gets a large mass $\sim \langle\phi_4\rangle$.
Upon integrating these fields out, the leading term in the 
effective superpotential is
\beq
W_{eff}\supseteq -\frac{\lambda_2^2}{4\lambda_1\langle\phi_4\rangle}\,
\phi_1^4.
\eeq
This mechanism produces operators with insertions
of $1/\langle\phi_V\rangle$ of the form
\beq
W_{eff}\supseteq \frac{M_*^a}{\langle\phi_V\rangle^a}\,
\frac{1}{M_*^{d-3}}\,\mathcal{O}^{(d)}(\phi_i).
\eeq
The possible charges of the operators $\mathcal{O}^{(d)}(\phi_i)$ 
are $+Q_V,+2Q_V,\ldots$.

\item {\it Supersymmetry breaking insertions.}
Supersymmetry breaking terms can transfer K\"ahler potential 
terms to the effective superpotential. 
This last mechanism can be schematically represented by
\beq
\int d^4\theta\;\frac{X^{\dagger}}{M_{*}}\,
\frac{1}{M_*^{a+b+d-2}}\,\phi_V^a\phantom{\!\!(}
{\phi_V^{\dagger}}\phantom{)\!\!\!}^b\,\mathcal{O}^{(d)}
\longrightarrow
\int d^2\theta\;\frac{F_X^{\dagger}}{M_{*}^2}\,
\frac{\langle\phi_V\rangle^a\langle\phi_V^{\dagger}\rangle^b}{M_*^{a+b}}\,
\frac{1}{M_*^{d-3}}\,\mathcal{O}^{(d)}
\label{susy breaking insertion}
\eeq
This mechanism always involves the supersymmetry breaking scale 
through the combination $\tilde m=F_X/M_{*}$, as well as possible
insertions of both $\langle \phi_V\rangle/M_*$ and 
$\langle\phi_V^{\dagger}\rangle/M_*$.  
\end{enumerate}

  As expected, at the level of supersymmetry breaking,
all operators consistent with the full $Z_{Q_V}$ symmetry
are allowed.  However, in the supersymmetric limit ($\tilde m\to 0$) 
operators with total charge equal to $-n\,Q_V$, where $n$ is a positive
integer, are generated by holomorphic insertions of
$\langle\phi_V\rangle$ (mechanism~1), while operators with total
charge $+n\,Q_V$ arise from integrating out holomorphic fields
leading to inverse-holomorphic insertions of $1/\langle\phi_V\rangle$ 
(mechanism~2).  
The coefficients of two operators with the same dimension
but opposite charge can therefore be very different.\footnote{
More generally, these mechanisms can operate simultaneously
leading to operator coefficients with powers of $\langle\phi_V\rangle$
in both the numerator and the denominator.  This does not change
the power counting or operator hierarchies discussed here.}  
If $\langle\phi_V\rangle \ll M_*$, operators generated by mechanism~2 
are potentially much larger than those from mechanism~1.  This is the 
essential difference between $Z_{Q_V}$ and $Z_{Q_V}^{hol}$.
 
  The distinction between $Z_{Q_V}$ and $\zhol_{Q_V}$ 
remains significant if there is a hierarchy 
$\tilde{m} \ll \left<\phi_V\right>$.  In this case the operators
containing insertions of $\tilde{m}$ are typically extremely suppressed
relative to those generated by mechanisms~1~and~2 above.
There is, however, one important exception.  The operators generated
by mechanism~2 depend on which fields in the full theory get
large masses due to the VEV.  If a particular operator with charge
$+nQ_V$ does not arise in the effective superpotential
due to mechanism~2, the operator will only appear due to mechanism~3.  
On the other hand, if the superpotential of the full theory is 
completely generic, mechanism~1 will generate every possible operator 
with charge $-nQ_V$ consistent with the other symmetries of the theory.


\subsection{Several VEVs: flat directions}

The $D$-term potential of a supersymmetric abelian gauge theory is
\beq
V_D=\frac{g^2}{2}D^2,~~{\rm where}~~D=\sum_i Q_i |\phi_i|^2+\xi.
\eeq
For simplicity we assume $\xi=0$, although our generic results do
not depend on this choice.  An anomaly-free theory must have charges of 
either sign.  Thus, whether or not $\xi$ is present, there is always 
a supersymmetric minimum of $V_D$ in which two fields with opposite-sign
charges develop VEVs.

  We shall focus on the case of two fields $\phi_a$ and $\phi_{-b}$
with charges $a$ and $-b$ obtaining large VEVs.  In practice, this
can occur if some fields have tachyonic soft mass squareds.
The $D$-term potential cancels provided their expectation values satisfy
\beq
b|\langle\phi_a\rangle|^2 = a|\langle\phi_{-b}\rangle|^2.
\eeq
Since the $D$-term potential is completely flat along this direction,
no particular value of $\langle\phi_a\rangle$ is favored.
This degeneracy is lifted and the VEVs are fixed by superpotential 
and supersymmetry breaking operators.  If the superpotential operators 
are small, either because they are higher dimensional or if they
have tiny couplings, the potential remains \emph{almost flat} along 
this direction, and the field VEV can be very large compared to the 
scale of supersymmetry breaking. 

  The $U(1)$ symmetry is broken along the almost flat direction by the VEVs
of $\phi_a$ and $\phi_{-b}$.  The residual symmetry is $Z_{(a,-b)}$  where
$(a,-b)$ is the greatest common divisor of $a$ and $b$.  If $a$ and $b$
are relatively prime numbers, there is no residual symmetry at all.
Even so, in a supersymmetric theory there is additional information
to be had.  To emphasize this point, we indicate the symmetry 
breaking pattern as
\beq
U(1)\stackrel{\langle\phi_{a}\rangle,\langle\phi_{-b}\rangle}
\longrightarrow \zhol_{(a,-b)}
~~({\rm supersymmetric}).
\eeq
The distinguishing feature between $\zhol_{(a,-b)}$ and $Z_{(a,-b)}$
are the relative sizes of operators appearing in the effective theory.

  To describe the effective theory below the $U(1)$ breaking scale,
it is helpful to write $\phi_a$ and $\phi_{-b}$ in the 
form~\cite{Poppitz:1994tx}
\beq
\phi_a = \sqrt{{b}}\,\tau\,e^{-ia\Omega},~~~~~
\phi_{-b} = \sqrt{{a}}\,\tau\,e^{ib\Omega},
\label{supergauge}
\eeq
where $\tau$ and $\Omega$ are chiral superfields.  The phase 
$\Omega$ can be gauged away completely, in which case the superfield 
degree of freedom is transferred to the $U(1)$ gauge multiplet which 
is integrated out.  The degrees of freedom associated with $\tau$ 
describe excitations along the flat direction.  To see why, note that all 
$D$-flat directions of condensing fields can be parametrized by the 
gauge invariant chiral polynomials of these 
fields~\cite{Affleck:1984xz,Luty:1995sd}.
In the present case, the only possibility is 
\beq
T=\phi_a^b\phi^a_{-b} = \sqrt{{a^bb^a}}\,\tau^{a+b},
\eeq
which is clearly in one-to-one correspondence with $\tau$.

  The operators in the low-energy superpotential are formed much
like in the case of a single VEV.  Instead of replacing $\phi_a$
and $\phi_{-b}$ with their expectation values, however, they
are replaced by their expressions from Eq.~(\ref{supergauge}) 
with $\Omega \to 0$.  The excitations of $\tau$ around its expectation 
value are light, and are thus still present in the effective theory.  
Integrating out heavy fields whose masses are proportional to the VEVs will
also generate operators with powers of $\phi_a$ and $\phi_{-b}$
in denominators.  The possible superpotential operators are therefore
\beq
\frac{\phi_a^s\phi_{-b}^t}{M_*^{d+s+t-3}}\;\mathcal{O}^{(d,Q)},
~~~~~~~Q = -sa+tb,\label{w-ops}
\eeq
where $\mathcal{O}^{(d,Q)}$ refers to an operator of dimension $d$ 
and charge $Q$, $s$ and $t$ are (possibly negative) integers,
and $\phi_a$ and $\phi_{-b}$ are to be expressed in terms of $\tau$.
In addition to these operators there are contributions
from supersymmetry breaking, but they are generally subleading.
The operator hierarchies due to $\zhol_{(a,-b)}$ are best understood
through an illustration.  Below, we derive a model of the $\mu$ term
based on these considerations.

\section{A small supersymmetric $\mu$ term}
\label{muterm}

  The operator hierarchy implied by a $\zhol_{(a,-b)}$ symmetry
together with an almost flat direction provides a mechanism for 
generating a naturally weak-scale value of the supersymmetric 
$\mu$-term.  Let $a$ and $b$ be relatively 
prime integers, and suppose the expectation values of the fields $\phi_a$ and 
$\phi_{-b}$ break a $U(1)_x$ gauge symmetry under which the 
$H_u\ccdot H_d$ superfield bilinear has charge $+1$.  The dominant
superpotential terms in the full theory are
\beq
W_{full} = \lambda_1\,\left(\frac{\phi_a^{q_a}\phi_{-b}^{q_b}}
{M_*^{q_a+q_b-1}}\right)\,H_u\ccdot H_d 
+ \lambda_2\,\left(\frac{\phi_a^b\phi_{-b}^a}{M_*^{a+b-3}}\right),
\label{weff}
\eeq 
where $M_* \sim M_{GUT}$ or $M_{\rm Pl}$ is an ultraviolet
cutoff scale, and $q_a$ and $q_b$ are the smallest positive integers such that
$q_aa-q_bb = -1$.  These superpotential operators break $D$-flatness
and drive the VEVs to zero.  To ensure non-zero expectation values,
we include tachyonic soft masses
\beq
\mathscr{L}_{soft} = m_a^2|\phi_a|^2 + m_b^2|\phi_{-b}|^2.
\eeq
Since the stabilizing effects of $F$-terms in the potential are
suppressed by powers of $M_*$, the vacuum expectation values
of $\phi_a$ and $\phi_{-b}$ are much larger than the soft masses.
A similar structure of the full potential for a $\mu$-term solution 
can be found in Ref.~\cite{example mu solutions}.

  Writing $\phia$ and $\phib$ in terms of $\tau$, the leading terms in
the scalar potential for $\tau$ are\footnote{We show in the appendix
that the K\"ahler potential for $\tau$ in the effective theory
is canonical up to small corrections.  Thus, the K\"ahler potential
does not play an important role in the bosonic potential for $\tau$.}
\beq
V_{\tau} = -\tilde{m}^2|\tau|^2 + \tilde{\lambda}_2M_*^4
\left(\frac{|\tau|}{M_*}\right)^{2(a+b)-2}
\label{taupot}
\eeq
where $\tilde m^2$ and $\tilde \lambda_2$ are obtained 
straightforwardly from the full potential. 
The scalar potential is minimized when
\beq
\langle|\tau|\rangle \sim M_*\left(\frac{\tilde m}{M_*}\right)^{1/(a+b-2)}.
\label{tauvev}
\eeq
So far, the potential depends only on the modulus of $\tau$.  A parametrically
important contribution to the potential that fixes the phase
of $\tau$ is the supersymmetry breaking operator
\beq
\int d^2\theta\;\frac{X}{M_*}\,\frac{\phi_a^b\phi_{-b}^a}{M_*^{a+b}}\,M_*^3
\longrightarrow \tilde{m}M_*^3\left(\frac{\tau}{M_*}\right)^{a+b}.
\label{aterm}
\eeq

  The solution for $\langle\tau\rangle$ implies an effective $\mu$ term of size
\beq
\mu_{eff}\sim M_*\left( \frac{\tilde m}{M_*}\right)^{\frac{q_a+q_b}{a+b-2}}.
\label{mueff equation}
\eeq
Clearly, not all choices of $a$ and $b$ will work since
$(q_a+q_b)/(a+b-2)=1$ is needed to get $\mu_{eff}\sim \tilde{m}$.
A general solution that guarantees this relation for any values of
$\tilde{m} \ll M_*$ is 
\beq
(a,-b)=(n+1,-n)~~~{\rm implying}~~~q_a=n-1,~q_b=n~~\Rightarrow
\frac{q_a+q_b}{a+b-2}=1,
\eeq
where $n$ is a positive integer.  

  We note that different choices 
of $(a,-b)$ can obtain different hierarchies of the $\mu$ term with 
respect to the supersymmetry breaking scale $\tilde m$.  This could be 
of importance for model-building in split supersymmetry~\cite{split susy}, 
when the $\mu$ term is desired to be much smaller 
than  $\tilde m$ in order for gauge coupling unification to work out.
For example, choosing parameters such that the exponent in
Eq.~(\ref{mueff equation}) is greater than 1 gives
\beq
\frac{q_a+q_b}{a+b-2}=1+\Delta, ~~\Rightarrow \mu_{eff}=\tilde m 
\left( \frac{\tilde m}{M_*}\right)^\Delta.
\eeq
The value of $\Delta$ can then be tuned, from the 
model-building perspective, to suppress the $\mu$ term compared to the
typical superpartner mass of $\tilde m$.

  In passing to the effective theory, we must verify that corrections
involving inverse powers of $\phi_a$ and $\phi_{-b}$ do not generate
a $\mu_{eff}$ larger than the one we have found.  For the charges 
$a = n+1$ and $b=n$, the only dangerous gauge-invariant combination is 
\beq
\frac{1}{\phi_{n+1}\phi_{-n}}\,H_u\ccdot H_d.
\eeq
For this to be an operator in the superpotential, three powers of 
$M_*$ or $\tilde{m}$ are needed to make up the dimension.  
Since positive powers of $M_*$ cannot arise from integrating out 
at scale $|\tau|\lesssim M_*$, there are no large corrections to $\mu_{eff}$.
Similarly, any contribution from supersymmetry breaking must come
in with a power of $\tilde{m}$, and is therefore sub-leading as well.
Thus, we see that the strong correlation between the charge and dimension
of superpotential operators allowed by $\zhol_{(a,-b)}$ leads to 
a naturally small effective $\mu$ term.

  As it stands, the model is potentially anomalous with respect to
the new $U(1)_x$ gauge symmetry.  To avoid anomalies, new exotic
matter is typically required and this can disrupt gauge unification.
Let us assume that the $U(1)_x$ charges of
the MSSM fields are family universal, and are consistent with an
embedding in $SU(5)$: $q=u=e$ and $d=l$.  For the usual MSSM 
superpotential operators to be gauge invariant, 
with $(h_u+h_d)=1$, we must have
\beq
q = -\frac{1}{2}h_u,~~~~~~~l = \frac{3}{2}h_u-1, 
\label{su5}
\eeq
Suppose we add to the model two pairs of doublets,
\beq
D_1\oplus\overline{D}_1 = ({\bf 1},{\bf 2},\frac{1}{2})_{q_1}\oplus
({\bf 1},{\bf 2},-\frac{1}{2})_{\bar{q}_1},~~~~~
D_2\oplus\overline{D}_2 = ({\bf1},{\bf 2},\frac{1}{2})_{q_2}\oplus
({\bf 1},{\bf 2},-\frac{1}{2})_{\bar{q}_2},
\eeq
with $U(1)_x$ charges such that $q_1+\bar{q_1} = -(n+1)$ and $q_2+\bar{q}_2=n$.
Since the doublets are vector-like under the SM gauge group, they will not
induce any pure SM anomalies.  The quantum numbers of these doublets
also imply that the mixed $SU(3)_c^2U(1)_x$, $SU(2)_L^2U(1)_x$, 
and $U(1)_Y^2U(1)_x$ anomaly conditions all have the form
\beq
X_i-n_g =0,~~~~~i=1,2,3
\eeq
where the $X_i$ represent the contributions from SM exotics other than 
$D_1$, $\overline{D}_1$, $D_2$, and $\overline{D}_2$.  All three anomaly 
conditions can be satisfied by including complete vector-like 
(with respect to the SM) $SU(5)$ multiplets, each of which automatically 
generates $X_1=X_2=X_3$ in our normalization.
Furthermore, such multiplets do not contribute at all to the 
$U(1)_YU(1)_X^2$ mixed anomaly, for which the cancellation condition is  
\beq
0 = (2h_u-1)-(q_1-\bar{q}_1)\,(n+1)+(q_2-\bar{q}_2)\,n.
\label{y1x2}
\eeq
One possible solution is $q_1=\bar{q}_1$, $q_2=\bar{q}_2$, and $h_u=1/2$.
Thus, by adding two pairs of doublets and some number of complete
$SU(5)$ multiplets, it is possible to cancel all the SM-$U(1)_x$
mixed anomalies in the model.  The remaining $U(1)_x^3$ 
and gravity-$U(1)_x$ anomalies can be eliminated by including 
SM gauge singlets (see, \emph{e.g.}, \cite{Batra:2005rh,Morrissey:2005uz}).

  Our solution to the anomaly constraints can also be consistent
with gauge unification.  In this regard, only the two pairs of doublets
pose a threat.  However, given their charges they can obtain large masses
when $\phi_{n+1}$ and $\phi_{-n}$ condense from the superpotential operators
\beq
W\supset \phi_{n+1}\bar{D}_1D_1 + \phi_{-n}\bar{D}_2D_2.
\eeq 
For very large VEVs, within a couple of orders magnitude of $10^{16}$~GeV,
the doublets will be very heavy, they will only slightly disrupt 
the running of the gauge couplings, and unification will be preserved.  
This mechanism is attractive because it circumvents some of the difficulties 
associated with more common $U(1)_x$ solutions to the 
$\mu$ problem with respect to unification~\cite{Morrissey:2005uz}.


\section{Thermal inflation}
\label{therminf}

  The extremely shallow potentials that arise naturally from breaking 
a supersymmetric $U(1)$ gauge symmetry can also play an
important role in the early universe.  When a potential is almost flat, 
thermal corrections often induce a metastable false vacuum.  
For a system stuck in such a vacuum, the excess vacuum energy may 
come to dominate the energy density of the universe giving rise to 
a period of late-time \emph{thermal inflation}~\cite{Lyth:1995ka,Adams:1997de}.
In the present section we show how this scenario is realized within 
the model considered above.

  The effect of thermal corrections on the potential depends on
the temperature relative to the zero-temperature expectation value,
$\tau_0\equiv \langle \tau\rangle_{T=0}$~\cite{Bertolami:1987xb}.  
At very high temperature, $T\gg \tau_0$, the exotic $U(1)_x$ gauge 
bosons and gauginos are abundant in the plasma, and they induce 
positive soft squared masses for $\phi_a$ and $\phi_{-b}$ of order 
$g_x^2T^2 \gg \tilde{m}^2$.  
In this case, the unique minimum of the finite-temperature effective 
potential lies at the origin, where the $U(1)_x$ gauge symmetry is unbroken.

\begin{figure}[htb]
\centerline{
        \includegraphics[width=0.45\textwidth]{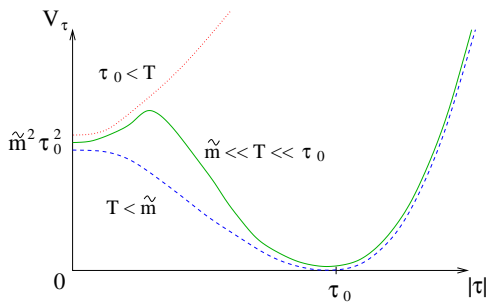}}
        \caption{An illustration of the temperature-corrected scalar potential.
The red (dotted) line shows the potential at very high temperatures 
$T\gg \tau_0$, while the green (solid) line corresponds to intermediate
temperatures in the range $\tilde{m}\ll T \ll \tau_0$, and the lower blue 
(dashed) line shows the potential for $T\ll \tilde{m}$.}
\label{tpotential}        
\end{figure}

  At temperatures smaller than $\tau_0$, but still much larger
than $\tilde{m}$, the potential has two minima~\cite{Bertolami:1987xb}.  
For small field values, $|\tau| \ll \tau_0$, the gauge bosons are light 
and they again generate soft squared masses of order $g_x^2T^2$.  
These thermal corrections induce a local minimum at the origin.  
Conversely, for large field values, $|\tau| \gg T$, 
the gauge bosons and gauginos are very heavy and the thermal 
corrections they induce are therefore Boltzmann-suppressed.  
Since the couplings of the $\tau$ excitations
to other fields are suppressed by powers of $\tau/M_*$, the effective 
potential is only slightly modified for these large values of $|\tau|$,
and a minimum near $|\tau| = \tau_0$ persists.  For $T\ll \tau_0$, this is
the global minimum of the potential.  The potential is illustrated
in Fig.~\ref{tpotential}.

  The cosmological effects of this potential are determined by whether or not 
$\tau$ is trapped in the local vacuum at $\tau = 0$ after primordial
inflation.  This will almost certainly be the case if the reheating 
temperature after inflation exceeds $\tau_0$.  Even for reheating temperatures
below $\tau_0$, the $\tau$ field may be trapped at the origin by the
``Hubble mass'' operator, $H^2|\tau|^2$, provided it is generated with 
a positive sign~\cite{Dine:1995uk}.  If the field is not trapped at the
origin, it will behave as a moduli field, and will be cosmologically 
dangerous if it decays after the onset of nucleosynthesis.

  Let us assume that the $\tau$ field becomes trapped at the origin.  
The tunneling rate from this local minimum to the true vacuum 
is typically extremely small for $T > \tilde{m}$~\cite{Yamamoto:1985rd}, 
so this minimum is metastable until $T\sim\tilde{m}$.  The vacuum energy 
of the false vacuum is of order $\tilde{m}^2\tau_0^2$ compared to the value 
at the global minimum.  If the universe is initially radiation-dominated,
then as the temperature cools below $T \sim \sqrt{\tilde{m}\tau_0}$ 
the excess vacuum energy becomes the dominant component of the total 
energy density, and the universe begins to inflate.  
The Hubble rate during this era is 
\beq
H = H_0 = \sqrt{\frac{8\pi}{3}}\frac{\tilde{m}\tau_0}{M_{\rm Pl}},
\eeq
which determines the expansion rate, $a(t) = a(t_0)e^{H_0(t-t_0)}$.
The exponential expansion ceases when the temperature falls below 
$T\sim \tilde{m}$ yielding a total number of $e$-foldings 
\beq
N_e \simeq \ln\left(\frac{\sqrt{\tilde{m}\tau_0}}{\tilde{m}}\right)
= \frac{1}{2}\ln\left(\frac{\tau_0}{\tilde{m}}\right).
\eeq
This number is of order $10$ for $\tilde{m} \sim 10^3$~GeV and
$\tau_0\sim 10^{12}$~GeV.  Such a small number of $e$-foldings
is not enough to disrupt the density perturbation induced by primordial
inflation~\cite{Randall:1994fr}.  The amount of inflation will be somewhat
less if the universe is matter dominated before thermal inflation.
For example, a moduli field with a Planck scale VEV and a shallow 
potential with curvature of order $\tilde{m}$ will dominate the energy
density of the universe once the temperature falls below 
$T\sim \sqrt{\tilde{m}M_{\rm Pl}}$.  This postpones the start of thermal
inflation, reducing the temperature at which inflation begins by a factor of 
$(\tau_0/M_{\rm Pl})^{1/6}$~\cite{Lyth:1995ka}, and decreasing 
the number of $e$-foldings by an amount $\ln(\tau_0/M_{\rm Pl})/6$.

  Once $T$ falls below $\tilde{m}$, the $\tau$ field rolls down
the potential towards the global minimum and begins to oscillate.
The oscillations dominate the total energy density until the 
$\tau$ field decays away.  Assuming the coupling between $\tau$ and the
Higgs fields to be of the form of Eq.~(\ref{weff}), we estimate this 
decay rate to be 
\beq
\Gamma_{\tau} = \gamma\frac{\tilde{m}^3}{\tau_0^2},
\eeq
where $\gamma$ is a dimensionless constant less than or on the order
of unity.  If the products of this decay thermalize rapidly,
the reheating temperature after the decay is~\cite{kolbturner}
\bea
T_{RH} &\simeq& \left(\frac{3}{\pi^3g_*}\right)^{1/4}
(M_{\rm Pl}\,\Gamma_{\tau})^{1/2}\\
&\simeq& 3~\mbox{GeV}\;\left(\frac{10}{g_*}\right)^{1/4}
\left(\frac{\tilde{m}}{1\mbox{TeV}}\right)^{3/2}
\left(\frac{10^{13}\mbox{GeV}}{\tau_0}\right)\,\gamma^{1/2}.\nonumber
\eea
The reheating temperature must be greater than about $5$~MeV to preserve
the successful predictions of nucleosynthesis~\cite{Hannestad:2004px}, 
and this implies an upper bound on $\tau_0$ on the order of $10^{16}$~GeV
(for $\gamma = 1$ and $\tilde{m}=1$~TeV).  For comparison, if we set 
$M_* = M_{\rm Pl}/\sqrt{8\pi}$ in Eq.~(\ref{tauvev}), 
we find $\tau_0\lsim 2\times 10^{13},~2\times 10^{15},~5\times10^{16}$~GeV 
for $(a+b) = 5,~7,~9$ where $a$ and $b$ are the powers in Eq.~(\ref{weff}).  
 
  Thermal inflation provides a mechanism to reduce the density of
unwanted relics.  In addition to dilution by the inflationary expansion, 
a decoupled relic is diluted even further by the entropy released when 
the $\tau$'s decay by a factor of order $\sqrt{\tilde m^5M_{\rm Pl}/\tau_0^6}$.
This dilution factor may even be needed to reduce the relic abundance
of late-decaying gravitinos and moduli that could disrupt big bang
nucleosynthesis, or of other particles with an overly large energy
density at late times~\cite{dilute relics}.
For gravitinos and moduli, the dilution factor in our model is 
sufficient to lower their abundance to an acceptable level provided 
$\tau_0 \gtrsim 10^{10}$~GeV~\cite{Lyth:1995ka}.
Unfortunately, desirable relics such as a dark matter particle or
a baryon asymmetry will also be diluted.  The extent to which they
are regenerated depends on $T_{RH}$, as well as the details of 
the $\tau$ decay.  Even for very low reheating temperatures,
well below 1~GeV, dark matter LSP's can be created non-thermally in 
the decays of the $\tau$~\cite{Gelmini:2006pw}, leading to a nonthermal
dark matter candidate.  The baryon asymmetry is more difficult to 
explain within this scenario, but it might be generated through new
dynamics associated with the flat direction~\cite{Lazarides:1985ja,
Stewart:1996ai}.



\section{Conclusions}
\label{concl}

  We have examined the operator hierarchies that emerge
from the spontaneous breakdown of a $U(1)$ gauge symmetry
in a supersymmetric theory.  The constraints induced by holomorphy
lead to large hierarchies between operators with the same mass dimension
but different charges.  We have made use of these hierarchies
to construct a naturally small supersymmetric $\mu$ term,
as well as a simple realization of thermal inflation.  The solution
to the $\mu$ term presented here lacks some of the challenges
of the more common approach of employing the vacuum
expectation value of a singlet field $S$ in the superpotential 
$\lambda SH_u\ccdot H_d$ to form 
$\mu_{eff}=\lambda\langle S\rangle$~\cite{Morrissey:2005uz}. 
Furthermore, the cosmology of thermal inflation, which is a natural 
byproduct of the $Z_{(a,-b)}^{hol}$ solution to the $\mu$ term presented 
in this work, allows for large suppressions of unwanted relics, 
such as late-decaying gravitino and moduli fields, while
simultaneously allowing for the existence of a good cold 
dark matter candidate.

\section*{Acknowledgements}
We thank P.C.~Argyres and S.P.~Martin for helpful conversations. 
This work is supported in part by the Department of Energy 
and the Michigan Center for Theoretical Physics~(MCTP).


\appendix

\section*{Appendix: Integrating out heavy superfields}

  In this appendix we describe the process of integrating out fields that 
become heavy upon the spontaneous breaking of a $U(1)$ gauge symmetry.
Along the way, we provide evidence for our claim that the operators
in the resulting low-energy effective superpotential arise from the
three mechanisms we described in the text:
holomorphic insertions, inverse-holomorphic insertions, 
and supersymmetry breaking insertions.
Throughout the analysis, we assume that the symmetry breaking
VEVs are much larger than the scale of supersymmetry breaking, and that
the superpotential is small in units of the VEV. (For example,
this is the case if the superpotential contains only higher 
dimensional operators.)  If this condition holds true,
the directions in field space that would be flat in the absence
of a superpotential or supersymmetry breaking remain almost flat
after the inclusion of these effects.  

  When there are almost flat directions, it is convenient to think of 
the process of forming the effective theory below the symmetry 
breaking scale as a three-step process.  The first step consists 
of parametrizing the $D$-flat directions, and integrating 
out the vector multiplet and the fields orthogonal to the flat directions
at an arbitrary point well out in the moduli space.
The second step is to integrate out fields that do not condense, 
but that develop large supersymmetric masses as a result of the 
symmetry breaking.  Thirdly, supersymmetry breaking is included as
a small perturbation.  For this procedure to be self-consistent,
the VEVs of the moduli fields must be much larger than the supersymmetry
breaking terms.

\bigskip
\noindent
{\Large \it Vector multiplets}

  Consider the case of two chiral superfields, 
$\phi_a$ and $\phi_{-b}$, obtaining large VEVs and a collection of other 
fields $\phi_i$ that do not.  The fields $\phi_{a}$ and $\phi_{-b}$ have 
charges $a$ and $-b$ respectively, and break the $U(1)$ symmetry when they 
condense. The leading terms in the K\"ahler potential are
\beq
\int d^4\theta\;{K} = 
\int d^4\theta\;\left({\phia}^{\dagger}e^{a\mathscr{V}}\phia 
+ {{\phi}_{-b}^{\dagger}}
e^{-b\mathscr{V}}\phib + {\phi}_i^{\dagger}e^{q_i\mathscr{V}}
e^{\mathscr{V}_{ew}}\phi_i\right).
\eeq
Here, we have allowed for the possibility that the $\phi_i$ also 
transform under another gauge group.

  Now suppose $\phia$ and $\phib$ develop large expectation values,
but the $\phi_i$ do not.  Following~\cite{Poppitz:1994tx}, we parametrize 
these fields as
\bea
\phia &=& \sqrt{b}\,e^{-ia\Omega}\,\tau\label{gauge}\\
\phib &=& \sqrt{a}\,e^{ib\Omega}\,\tau\nnmb\\
\phi_i &=& e^{-iQ_i\Omega}\tilde{\phi}_i,\nnmb
\eea
where $\Omega$, $\tau$, and $\tilde{\phi}_i$ are chiral superfields.
To maintain this parametrization under $U(1)$ supergauge transformations,
we take $\Omega$ to transform by a shift, and $\tau$ and $\tilde{\phi}_q$ 
to be invariant.  Note that $\Omega$ can be gauged away completely.
The motivation for this form is that 
$\tau$ is in one-to-one correspondence with the unique gauge invariant 
polynomial that we can make from $\phia$ and $\phib$,
\beq
\phia^b\phib^a = \sqrt{a^ab^b} \tau^{a+b}.
\eeq
Thus, $\tau$ parametrizes the $ab$ flat direction~\cite{Luty:1995sd}.

  The vector multiplet can be integrated out using its superfield 
equation of motion,
\bea
0&=&\frac{d{K}}{d\mathscr{V}} + (D^2,\bar{D}^2~terms)\\
&\simeq& ab\,{\tau}^{\dagger}\tau\,e^{a[\mathscr{V}-i(\Omega
-\bar{\Omega}^{\dagger})]}
- ab\,{\tau}^{\dagger}\tau\,e^{-b[\mathscr{V}-i(\Omega
-\bar{\Omega}^{\dagger})]}
+ Q_i\,{\tilde{\phi}}_i^{\dagger}e^{\mathscr{V}_{ew}}\tilde{\phi}_i\,
e^{Q_i[\mathscr{V}-i(\Omega-{\Omega}^{\dagger})]}.\nnmb
\eea
Treating the $\phi_i$ as small and $\tau$ as large, the solution is
\beq
\mathscr{V} -i(\Omega-{\Omega}^{\dagger}) = 0 + \mathcal{O}
\left(\frac{{\tilde{\phi}}_i^{\dagger}e^{\mathscr{V}_{ew}}
\tilde{\phi}_i}{{\tau}^{\dagger}\tau}\right).
\eeq
Therefore the leading terms in the K\"ahler potential of the 
effective theory are
\beq
{K}_{eff} = (a+b)\,{\tau}^{\dagger}\tau + {\tilde{\phi}}_i^{\dagger}
e^{\mathscr{V}_{ew}}\tilde{\phi}_i.
\eeq
Up to a trivial rescaling and small corrections, the K\"ahler potential
for $\tau$ is canonical.  

  If the full theory also has a superpotential, then by gauge invariance 
it can only be a function of $\tau$ and $\tilde{\phi}_i$,
but not $\Omega$.  Thus, in passing to the effective theory, 
the procedure of integrating out the vector multiplet (and in the
process the gauge artifact $\Omega$) only affects the K\"ahler potential.  
The resulting superpotential is simply given by its expression in the
full theory with the replacements $\phi_{a,-b}\to \tau$ 
and $\phi_i\to\tilde{\phi}_i$.
This is the source of the holomorphic insertions described in the text.
The light field corresponding to the almost flat direction appears in
the effective theory by expanding $\tau$ about its VEV.

  The case of a single field obtaining a VEV can be treated in the same way.  
Now there is no flat direction, and the condensing field is 
eaten by the vector multiplet.  This is manifest if we express the
condensing field $\phi_V$ in the form
\beq
\phi_V = e^{-iQ_V\,\Omega}\,\langle\phi_V\rangle,
\eeq
where $\langle\phi_V\rangle = \xi/Q_V$~(see Eq.~(\ref{qvxi})),
and then make a supergauge transformation to remove $\Omega$.  
The equation of motion for the vector multiplet then gives
$\mathscr{V} = 0$, up to small corrections of order $|\phi_i|^2/
\langle|\phi_V|^2\rangle$.  By gauge invariance,
$\Omega$ cannot appear in the full superpotential, so in the 
effective superpotential $\phi_V$ is simply replaced by its 
expectation value.  There is no expansion about this value because there
is no flat direction.  

  When three or more fields develop large VEVs,
we can again use a similar technique.  However, in this case the
equation of motion for $\mathscr{V}$ is typically very complicated,
and the effective K\"ahler potential need not have the minimal form
found in the two VEV scenario.


\bigskip
\noindent
{\Large\it Chiral multiplets}

  The next step is to consider the effects of the VEVs on the
superpotential.  In the effective theory, the flat direction
fields such as $\tau$ are expanded about their VEVs.  By construction,
these fields have masses parametrically smaller than the VEVs.
However, the appearance of large expectation values can give rise to 
large supersymmetric masses for other fields that do not condense.
These heavy fields should also be integrated out.  We show here
that the integration out procedure can be performed in such a way
that the resulting effective superpotential will be holomorphic
in both the light fields and the parameters in the full superpotential,
up to higher derivatives and supersymmetry breaking~\cite{Dine:1994su}.
 
  Suppose only one chiral superfield $\Phi$ develops a large mass 
due to the VEVs.  The full superpotential must therefore be of the form
\beq
W_{full} = \frac{1}{2}M\Phi^2 - f(\Phi,\phi),
\eeq
where $M$ denotes the large mass, proportional to the VEV,
and $\phi$ refers to any field other than $\Phi$.  By assumption, 
$f(\Phi,\phi)$ contains no positive dimensional couplings unless 
they are much smaller than $M$.  
The equation of motion for $\Phi$ can be expressed in the form
\beq
\Phi = \frac{1}{M}\left[\frac{\del f(\Phi,\phi)}{\del\Phi} 
+ \frac{\bar{D}^2}{4}\frac{\del K}{\del\Phi}\right],
\label{phisol}
\eeq
where $K$ denotes the K\"ahler potential.
We can solve iteratively for $\Phi$ by replacing
$\Phi$ on the right hand side with this relation.  
Note that each repetition of this procedure always brings in an 
additional power of $1/M$, and thus the solution is expected
to converge rapidly.  Since inverse powers of $M$ appear in the solution
for $\Phi$, they will also appear in the effective superpotential.
This is the source of the inverse holomorphic insertions (mechanism~2)
described in the text.

  We claim that to any order in this procedure, the resulting
expression for $\Phi$ can be written in the form
\beq
\Phi = (holomorphic) + \frac{\bar{D}^2}{4}(maybe~non\!-\!holomorphic),
\label{form}
\eeq
where the first term is holomorphic in both the fields and all the 
superpotential parameters.  At lowest order in the $1/M$ expansion
we set $\Phi = 0$ on the right hand side of Eq.~(\ref{phisol}) and our
assertion is clearly satisfied.  If we assume that at the $n$-th order
our claim is true, then inserting Eq.~(\ref{phisol}) and expanding,
we see that $\Phi$ will have the form of Eq.~(\ref{form}) at the 
$(n+1)$-th order as well.  Thus, our assertion follows by induction.

  This result is enough to show that the effective superpotential
will be holomorphic in the couplings of the full superpotential, 
up to supersymmetry breaking and higher derivative operators.
Inserting the full solution for $\Phi$, in the form of Eq.~(\ref{form}),
into the superpotential we obtain holomorphic terms without derivatives,
as well as terms of the type
\beq
\int d^2\theta\;G(\phi)\left(\frac{-\bar{D}^2}{4}\right)H({\phi}^{\dagger}).
\eeq
These can be converted into K\"ahler potential terms using the fact 
up to a total derivative, $-\bar{D}^2/4$ is equivalent 
$d^2\bar{\theta}$~\cite{Wess:1992cp}.
Putting our solution for $\Phi$ into the K\"ahler potential, there
can also arise terms of the form
\beq
\int d^4\theta\;A(\phi)\left(-\frac{D^2}{4}\right)B(\phi)
= \int d^2\theta\;{A}(\phi)\del^2{B}(\phi),
\eeq
where we have made use of the identity 
$(\bar{D}^2D^2/16)\phi = \del^2\phi$ for any chiral superfield $\phi$.
These higher derivative operators can be non-holomorphic in the 
parameters of the full superpotential or the VEVs, but they are expected
to be negligible at low energies.  Soft supersymmetry breaking can also
be included in this procedure by treating the coefficients in the full
theory as constant superfields with non-zero auxiliary components.

  Up to one subtlety, it is not hard to generalize this result
to many large mass terms.  In this case, the full superpotential 
can be written in the form
\beq
W_{full} = \frac{1}{2}\Phi_i\mathcal{M}_{ij}\Phi_j - f(\Phi,\phi),
\label{wfull}
\eeq
where, again, $f(\Phi,\phi)$ contains no large couplings with 
positive mass dimension, and $i,j = 1,2,\ldots N$.
Without loss of generality, we may assume that the elements of 
$\mathcal{M}_{ij}$ are all either of order the large VEV scale $M$ or zero, 
and that for a given $i$ there is at least one value of $j$ for which 
$\mathcal{M}_{ij} \neq 0$.

  The equations of motion for the $\Phi_i$ now become
\beq
\mathcal{M}_{ij}\Phi_j = \frac{\del f}{\del\Phi_i} 
+ \frac{\bar{D}^2}{4}\frac{\del K}{\del\Phi_i}.
\eeq
If $\det(\mathcal{M}) \sim M^N$, the mass matrix has no small eigenvalues,
and we can take (holomorphic) linear combinations of the equations 
of motion such that we end up with 
\beq
\Phi_i = \frac{1}{\tilde{M}_i}\left[\tilde{f}'_i(\Phi,\phi) 
+ \frac{\bar{D}^2}{4}\tilde{K}'_i\right],
\eeq
where $\tilde{M}_i$ is a rational holomorphic function of the 
$\mathcal{M}_{ij}$ of order $M$ and has a well-defined spurious charge,
$\tilde{f}'_i$ is a linear combination of the $\del f/\del\Phi_j$,
and $\tilde{K}'_i$ is a linear combination of the $\del K/\del\Phi_k$.
This expression is holomorphic up to the $\bar{D}^2$ term, and therefore
our previous argument applies.

  This procedure breaks down when the matrix $\mathcal{M}$ has an eigenvalue
that is zero or much smaller than the large mass scale $M$.  This may be
due to a symmetry, or the result of an accidental cancellation.
Either way, this implies that at least one linear combination of the $\Phi_i$
is a light degree of freedom that should not be integrated out.
To identify these light states, we need only find the (approximate) null space
of $\mathcal{M}$.  The resulting null vectors will correspond to holomorphic
linear combinations of the $\Phi_i$ with well-defined spurious charges
that remain light in the effective theory.  By taking holomorphic linear
combinations of the fields, it is possible to form a new field basis
comprised of the null vectors and some other non-null combinations.
In this basis we can integrate out the non-massless fields and apply
our previous arguments.

  The main result of this section is that up to supersymmetry breaking
and higher derivative interactions, the procedure of integrating out
heavy chiral superfields yields an effective superpotential that is
holomorphic in the light fields as well as all the parameters 
(including the VEVs) present in the full superpotential.  
Non-holomorphic parameter dependences can only appear in higher-derivative
interactions in the effective superpotential or from supersymmetry breaking.
Eq.~(\ref{phisol}) also shows that the integration-out procedure 
can generate inverse-holomorphic insertions of the VEV (or large mass) 
in the effective superpotential.

  As a simple example of this process, consider a model with a single
heavy chiral field $\Phi$ of mass $M$ and minimal K\"ahler potential,
interacting with the light chiral fields $\phi$ through 
the superpotential~\cite{Cvetic:1998ef}
\beq
W = \frac{1}{2}\,M\Phi^2 + \Phi\,W_1(\phi)+W_0(\phi).
\eeq
The classical solution for $\Phi$ is
\beq
\Phi = -\left(1-\frac{\del^2}{{M}^{\dagger}M}\right)^{-1}
\left[\frac{1}{M}W_1(\phi)
+\frac{1}{{M}^{\dagger}M}\frac{\bar{D}^2}{4}{W}_1^{\dagger}({\phi}^{\dagger})
\right].
\eeq
Replacing $\Phi$ by this solution, the low energy effective action becomes
\beq
S_{eff}=\int d^4xd^2\theta d^2\bar \theta\; K_{eff}+\left[\int 
d^4xd^2\theta\; W_{eff}+ (h.c.)\right]
\eeq
where
\bea
K_{eff} & = & K_{light}+ {W}_1^{\dagger} \frac{1}{{M}^{\dagger}M}
\left( 1 -\frac{\del^2}{{M}^{\dagger}{M}}\right)^{-1}W_1, \\
W_{eff}& = & W_0(\phi)+ W_1\frac{1}{{M}} 
\left( 1 -\frac{\del^2}{{M}^{\dagger}{M}}\right)^{-1} W_1.\nnmb 
\eea

  The first two mechanisms described in the text 
are illustrated in the above expressions.  $W_0(\phi)$ contains, 
in general, holomorphic insertions that lead to couplings proportional 
to $\langle\tau\rangle/M_*$, where $M_*$ is the cutoff of the 
original high-energy theory. The second term in $W_{eff}$ contains 
inverse-holomorphic insertions of $1/{M}$, as well as higher-order, 
subleading corrections involving $\del^2/{M}^{\dagger}M$.
Supersymmetry breaking insertions may also included as small
perturbations to this picture.
Some supersymmetry breaking insertions can lead to terms in 
$W_0(\phi)$, as Eq.~(\ref{susy breaking insertion}) suggests, while 
others are most easily captured by the addition of a 
soft supersymmetry breaking Lagrangian contribution, ${\cal L}_{soft}$, 
outside of the $W_{eff}$ or $K_{eff}$ language.


\end{document}